# Room-temperature Optically Detected Magnetic Resonance of Telecom Single Photon Emitters in GaN

John J.H. Eng [1,2†], Zhengzhi Jiang [3,4†], Max Meunier [1], Abdullah Rasmita [1], Haoran Zhang [1], Yuzhe Yang [1], Feifei Zhou [1], Hongbing Cai [1,5], Zhaogang Dong [2], Jesús Zúñiga Pérez [1,6,7*], Weibo Gao [1,5*]

[1] *Division of Physics and Applied Physics, School of Physical and Mathematical Sciences, Nanyang Technological University, Singapore 637371, Singapore.*

[2] *Institute of Materials Research and Engineering (IMRE), Agency for Science, Technology and Research (A\*STAR), 2 Fusionopolis Way, Innovis #08-03, Singapore 138634, Republic of Singapore*

[3] *Joint School of National University of Singapore and Tianjin University, International Campus of Tianjin University, Binhai New City, Fuzhou 350207, P. R. China.*

[4] *Department of Chemistry, National University of Singapore, Singapore 117543, Singapore.*

[5] *The Photonics Institute and Centre for Disruptive Photonic Technologies, Nanyang Technological University, Singapore.*

[6] *Université Côte d'Azur, Centre National de la Recherche Scientifique (CNRS), Centre de Recherche sur l'Hétéro Epitaxie et ses Applications (CRHEA), Rue Bernard Gregory, 06560 Valbonne, France*

[7] *Majulab International Joint Research Unit UMI 3654, CNRS, Université Côte d'Azûr, Sorbonne Université, National University of Singapore, Nanyang Technological University, Singapore*

*Email: jesus.zuniga@ntu.edu.sg, wbgao@ntu.edu.sg*

**Abstract**

Solid-state defects susceptible of spin manipulation hold great promise for scalable quantum technology. To broaden their utility, operating at room temperature and emitting in the telecom wavelength range are desired, eliminating cryogenic requirements and leveraging existing optical fiber infrastructure for transmitting the quantum information. To that end, we report that telecom single photon emitters (SPEs) in gallium nitride (GaN) exhibit optically detected magnetic resonance (ODMR) at room temperature. The analysis of ODMR as a function of magnetic field orientation enables the determination of the orientation of the spin quantization axis with respect to the GaN crystalline lattice. The optical transitions dynamics are analyzed to gain further insight into the transition rates dominating ODMR. Our findings, coupled with GaN's mature fabrication technology, could facilitate the realization of scalable quantum technology.

**Introduction**

Optically active spins based on defects in solid-state materials could pave the way for scalable quantum technology [1,2] with applications ranging from quantum computing to quantum communication [3-5]. One prime candidate is nitrogen vacancy (NV) centers in diamond [6-8], boasting spin coherence times exceeding one second [9] as well as high fidelity read-out [10]. While, a quantum network based on NV centers has already been realized [11], to leverage the low attenuation of existing optical fiber infrastructure, emission in the telecommunication wavelength range would be beneficial. One strategy is to utilize frequency conversion through non-linear optical processes [12-14]. However, the integration of frequency converters could be technologically challenging and reduce overall system efficiency. Spin defects in silicon carbide (SiC), such as divacancies and SiC NV centers [15-18], as well as point defects in silicon (Si) [19] have also shown their capability and potential in quantum applications benefiting from their telecom emission.

Recently gallium nitride (GaN), a semiconductor material industrially commercialized for its electronic and optoelectronic applications [20-22] including transistors, lasers and light-emitting diodes, has gathered increasing research interest as a platform for quantum technologies. GaN thin films have been found to host bright single photon emitters (SPEs) emitting both in the visible [23-27] and in the telecom [28,29]. In 2024, GaN SPEs emitting in the visible have been found to exhibit optically detected magnetic resonance (ODMR), which given their wavelength could be a promising platform for quantum sensing applications [30]. Beyond these quantum sensing applications, and given that first integrated photonic realizations in GaN begin to be developed [31-34], combining telecom GaN SPE, integrated photonics and telecommunication fiber networks could open new avenues for room-temperature based quantum technologies.

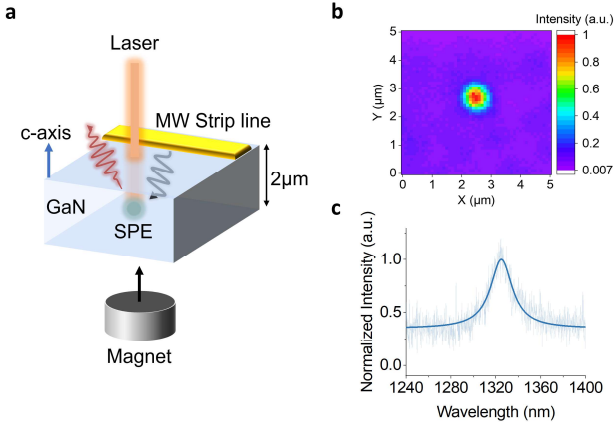

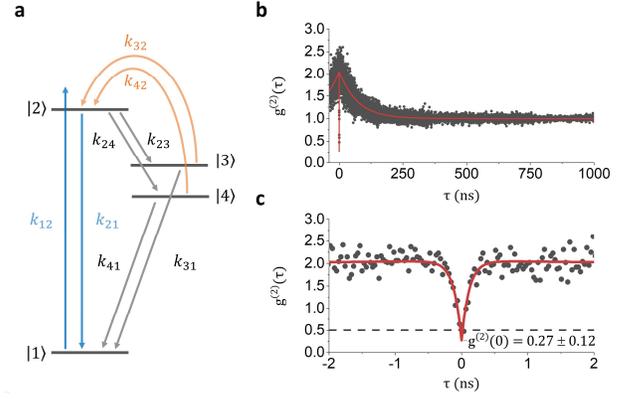

**Figure 1 | Typical optical properties of telecom SPEs. a,** Schematic of the confocal microscope setup for magneto-optical measurements. **b,** PL raster scan image of the emitter. **c,** PL spectrum of the emitter.

**Figure 2 | Model and second-order autocorrelation of the emitter. a,** Proposed 4-level energy system model. The blue arrows represent the optical transitions (excitation and recombination) involved in the single-photon emission, while the orange and black arrows represent transitions relevant to the overall singe-photon emission process. **b,** Continuous wave second-order autocorrelation $g^{(2)}(\tau)$ as a function of time delay $\tau$. The black scatter plot and the red solid line are the experimental data and its fitting line, respectively. **c,** Close-up image of the $g^{(2)}(\tau)$ in **b** around $\tau = 0$ with $g^{(2)}_{raw}(0) = 0.27 \pm 0.12$.

Here, for the first time, we report ODMR of telecom SPE in GaN at room temperature. Magnetic field-dependent ODMR is measured to extract the spin information, the value of zero-field splitting, and the gyromagnetic ratio. By analyzing the magnetic-field orientation dependence of the ODMR contrast, the spin quantization axis direction of the SPE with respect to the GaN lattice is established. Finally, to gain insight into the transition dynamics of the emitter including the ODMR process, we characterized the pump-probe (or time-gated) photoluminescence (PL) and proposed a possible energy diagram. Thus, our results provide valuable information concerning the nature of the defects at the origin of the telecommunication GaN SPE and open the door for practical and scalable quantum sensing and communication technologies based on spin-defects in GaN.

Fig. 1a shows a schematic of the home-built confocal microscope setup equipped with a movable magnet to characterize the magneto-optical properties of the telecommunication SPE. The sample consists of a non-intentionally doped 2μm-thick (0001) Ga-polar GaN thin film grown on c-plane sapphire by metalorganic chemical vapour deposition. The microwave field necessary to perform ODMR is applied through a lithographically deposited strip line on the top surface of the GaN layer. We employ PL raster scans in the area surrounding the strip line's edge to search for SPEs. The PL mapping (Fig. 1b) and spectrum (Fig. 1c) of the emitter were measured under 1mW off resonant (980nm) excitation. The spectrum shows that the emitter lies in the telecommunication wavelength range with an emission wavelength of ~1325 nm (i.e. in the O-band).

To verify the single photon emission nature of the emitter, we conducted second-order autocorrelation $g^{(2)}(\tau)$ measurements using the Hanbury-Brown Twiss (HBT) setup, with the laser power at 1mW (Fig. 2b) under ambient conditions. The $g^{(2)}_{raw}(\tau)$ curve is fitted with a general empirical model for a quantum emitter with multiple energy levels [35]:

$$g^{(2)}(\tau) = 1 - C_1 e^{-|\tau|/t_1} + \sum_{i=2}^{n} C_i e^{-|\tau|/t_i}, \quad (1)$$

where $C_1$ and $t_1$ are the antibunching amplitude and antibunching time, respectively, $C_i$ and $t_i$ for $i \geq 2$ are the bunching amplitudes and bunching times, respectively. An $N$-level system emitter should correspondingly have at least $N-1$ rates involved in the emission dynamics [35,36]. For this emitter, we found that the $g^{(2)}_{raw}(\tau)$ curve fits well with $n = 2$, suggesting the emitter is at least a 3-level energy system (Fig. 2a), consistent with previous results [28,29]. In particular, the value of $g^{(2)}_{raw}(0) = 0.27 \pm 0.12 < 0.5$ measured at 1mW without background correction, confirms its single photon emission. From the fitting of the experimental $g^{(2)}_{raw}(\tau)$ curve and numerical modelling of the rate equations, we associate $t_1 = 0.11 \pm 0.01$ns to the lifetime of the emitting state (state $|2\rangle$), while

$t_2 = 70.0 \pm 0.6$ns is related to the optical excitation process. This assignment is supported by the $g_{raw}^{(2)}(\tau)$ fitting parameters (using the general empirical model) at various pumping laser powers, where $t_1$ appears unaffected by the pumping laser power whereas $t_2$ is linearly dependent on pumping laser power. The fitting parameters of the $g_{raw}^{(2)}(\tau)$ at 4 different pumping laser powers as well as the theoretical model for the SPE $g^{(2)}(\tau)$ fitting via solving the coupled rate equations can be found in Table S1, Note 1 and Fig. S1 in [37].

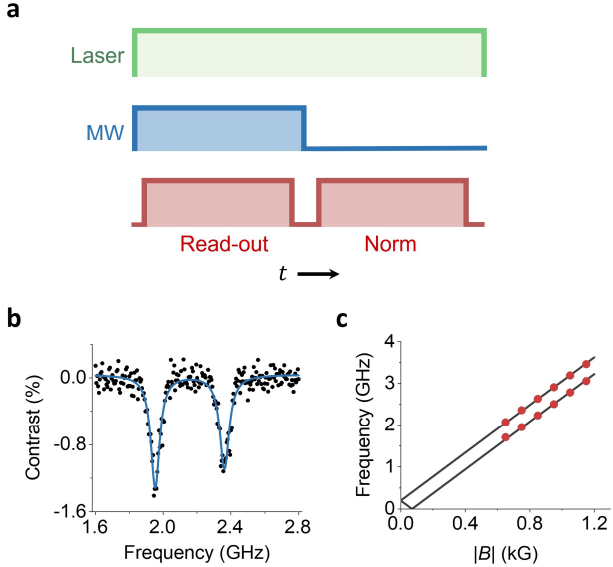

**Figure 3 | Magneto-optical properties of telecom SPE in GaN.**
**a**, Pulse sequence employed for CW-ODMR. **b**, CW-ODMR spectrum at $|\vec{B}| = 0.75$kG. The experimental points (black scatter points) are fitted with two Lorentzian functions (blue solid line). **c**, Magnetic flux density as a function of CW-ODMR resonant frequency. Red scatter plot and black solid lines are the experimental data and fitting lines, respectively. The experimental points are fitted with a simplified model.

To characterize the magneto-optical properties of the SPE, we performed continuous wave (CW)-ODMR measurement under ambient conditions with the measurement sequence shown in Fig. 3a. The contrast of ODMR is defined as: $(N_{read} - N_{norm})/(N_{read} + N_{norm})$. Here, $N_{read}$ and $N_{norm}$ are the photon counts during the read-out and normalization windows, respectively. During read-out window, microwaves are delivered by the strip line, while no microwaves are delivered during the normalization window. The measurements in Fig. 3b and c are conducted with the magnetic field vector maximally aligned to the spin quantization axis of the defect (see Fig. 4 and discussion below) by moving the magnet's stage in three orthogonal directions. See Note 2 and Fig. S2 in [37] for the calibration of the magnetic field amplitude and direction. The CW-ODMR spectrum presented in Fig. 3b, obtained at $|\vec{B}| = 0.75$kG and a microwave power of 40dBm, reveals two distinct transitions. This observation suggests the presence of a quantum system with a spin number $S = 1$, as indicated by the two transitions ($m_s = 0$ to $m_s = -1$ and $m_s = 0$ to $m_s = +1$).

To further examine the magnetic field-dependent spin resonance, depicted in Fig. 3c, we adjusted the strength of the magnetic field while ensuring that the magnetic field vector and spin quantization axis are aligned. We observe that the resonance depends linearly on the magnetic field. Here, we consider the simplest electronic spin Hamiltonian for an S=1 system, given by:

$$H = DS_z^2 + E(S_x^2 - S_y^2) + g\mu_B \vec{B} \cdot \vec{S} \qquad (2)$$

where $D$ and $E$ are the axial and transversal zero-field interaction parameters (so-called zero-field components), $S_{x,y,z}$ are the electronic spin operators, and $g$, $\mu_B$, and $\vec{B}$ are the electron g-factor, Bohr magneton, and applied magnetic field, respectively. Here, the $z$ axis is parallel to the defect spin quantization axis. The last term in the Hamiltonian describes the Zeeman effect on the emitter. Based on Eq. (2), we fit the resonance frequencies with the expression $\left|D \pm \sqrt{E^2 + (\gamma B_z)^2}\right|$, where the gyromagnetic ratio $\gamma = g\mu_B$. From the fitting, the axial zero-field splitting value $D$ is found to be $0.200 \pm 0.013$GHz, while the transverse zero-field splitting $E$ is zero. The gyromagnetic ratio $\gamma$ is fitted to be $28.55 \pm 0.14$GHz/T, close to 28 GHz/T of an electron spin.

We found that the SPE in Fig. 1 and 2 does not exhibit pulsed ODMR signal, nor it is observed from another SPE ($g_{raw}^{(2)}(0) = 0.07 \pm 0.22$) in the same sample displaying a similar magnetic field dependence (see Fig. S3 in [37]). The absence of pulsed ODMR could be attributed to the spin-dependent transitions occurring at the metastable state (i.e., states |3⟩ and/or |4⟩ in Fig. 2a), as suggested in a previous study [30], rather than at the ground or excited-states (i.e. states |1⟩ and |2⟩ in Fig. 2a).

Owing to the presence of ODMR signal, we could identify the spin quantization axis with respect to the GaN wurtzite lattice, and thereby provide essential information to identify the nature of the point-defect at the origin of the telecommunication SPE. Thus, to determine the orientation of the spin quantization axis of the SPE, we systematically moved the magnet along three Cartesian orthogonal directions to vary the angle of the magnetic field while maintaining the magnetic flux density at approximately 0.75kG (see Note 2 and Fig. S2 in [37] for the calibration and simulation of the magnetic field). Fig. 3a and b show the ODMR peak contrast as a function of magnetic field orientation, defined by the polar angles $\theta_x$ and $\theta_y$, that is, the projection angle of the magnetic field vector on the GaN's *c*-axis (Cartesian z axis) along the $x$ and $y$ axes, respectively, shown in Fig. 4c and d. The ODMR contrast should reduce when increasing the misalignment of the magnetic field vector and spin quantization axis. This can be explained by the augmented mixing of the eigenstates, similar to the results reported in [30]. The final values of $\theta_x$ and $\theta_y$ are $-23.3 \pm 1.6°$ and $-14.7 \pm 1.6°$, respectively. When translated to cylindrical polar coordinates, these angles correspond to a polar angle $\theta = 27 \pm 2°$ with respect to the *c*-axis and an azimuthal angle $\varphi = 1.3 \pm 1.1°$ with respect to the *a*-axis. Thus, the quantization axis does not correspond to any lattice vector of the GaN crystal [22], which suggests that the defect involves an interstitial atom.

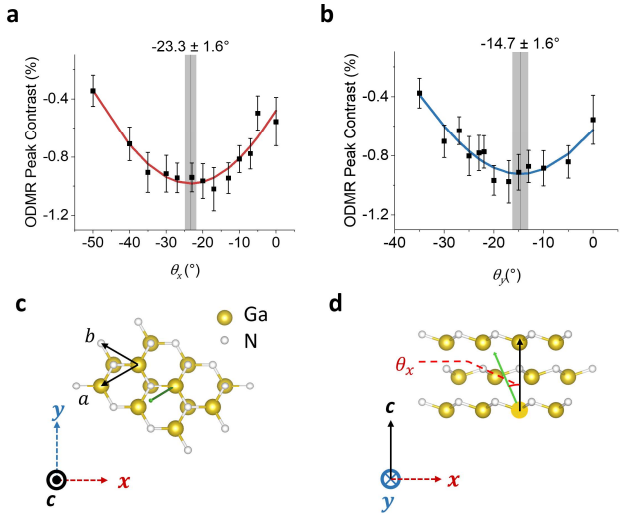

**Figure 4 | Angle dependent ODMR**. **a** and **b**, ODMR peak contrast with respect to the angles $\theta_x$ and $\theta_y$, which are the polar angles with respect to the GaN *c*-axis measured along $[11\bar{2}1]$ and $[10\bar{1}0]$ respectively. These polar angles $\theta_x$ and $\theta_y$ correspond to the angles between the magnetic field vector projected on the z (or *c*-axis of GaN) along the direction of the red $x$ and blue $y$ axes respectively as shown in **c** and **d**. The black square points are the experimental data with error bars acquired from the Lorentzian function fitting of the ODMR spectrum measured at multiple angles. The red and blue solid lines serve as visual guide in **a** and **b** respectively. The gray solid lines are the centre of the visual guidelines while the shaded gray areas are its error bars. **c** and **d**, Spin quantization axis (green arrow) of the telecom SPE with respect to the GaN's wurtzite lattice: views along the GaN [0001] and GaN $[10\bar{1}0]$ directions, respectively.

Interestingly, the orientation of the spin quantization axis, the electronic spin (S=1), and the amplitude of the peak ODMR contrast are similar to those of visible SPEs (defects in group I) reported in [30]. The differences are the emission wavelength and the SPE's photo-stability. Here, the telecommunication SPEs exhibiting ODMR do not suffer from noticeable photo-bleaching and blinking (for more than six months long period). Still, in previous studies, we also noticed that some telecommunication SPEs could show some photo-instability, especially at pumping powers larger than the saturation power [29]. Nonetheless, the many similarities between group I visible SPEs in [30] and our reported telecommunication SPEs could be due to them being the same type of defect, albeit with different charged states. Indeed, *ab initio* studies on antisites and vacancies complexes in GaN [38,39] have shown that charged states of the same defect species can emit both in the visible and telecommunication wavelengths. Interestingly, the zero-phonon lines (ZPL) in the *ab initio* studies are close to the reported emission spectra of the group I defects in the visible and in our report. Despite the similarities between theoretical calculations and experiments, there are essential differences. First, based on the angle dependent ODMR, both visible (group I) and telecommunication SPEs are likely to be related to interstitial defects, while antisites are substitutional defects thus, one should expect the quantization axis to be aligned with a lattice vector. Furthermore, the lifetime of the telecommunication SPEs reported in the current work are in the sub-nanosecond range, while the *ab initio* studies predict that the analyzed antistite defects emitting in the telecommunication range should exhibit radiative lifetimes on the order of 200ns. Alternatively, both SPEs could be completely different species of interstitial defect that coincides in the same quantization axis within several degrees. Thus, our work provides insightful quantitative information on the telecommunication SPEs that should motivate further theoretical calculations addressing specifically point defects emitting in the telecommunication wavelength range [40].

To gain further insight into the optical dynamics of the emitter, we implement a pump-probe experiment (or time gating technique), which consists of measuring time-resolved PL during a periodic sequence comprising two off-resonant excitation laser pulses with varying time delays $\tau$ between them (Fig. 5a). The 980nm laser pulses are controlled by an acousto-optic modulator (AOM) and the laser power is set to 1mW while operating in the CW mode, to ensure that the power during the illumination window is consistent with the $g^{(2)}(\tau)$ measurement. This approach is similar to the study of the emission dynamics of SPE in AlN [41] and contrasts with the typical pump-probe technique involving multiple different lasers. The resulting curve (Fig. 5b) shows the PL time trace at time delay $\tau = 1\mu s$ between each 5μs pulses. The first laser pulse can be associated to the "pump" pulse. As shown in Fig. 5b, the emitter reaches the steady-state condition within the pulse duration of 5μs. Immediately after the first pulse is switched off, the emitter should remain in the higher energy states (state $|2\rangle, |3\rangle,$ and $|4\rangle$) for some time before returning to thermal equilibrium. During this time, the second laser pulse (similar to the "probe" pulse) is introduced at a time delay $\tau$ after the first laser pulse is switched off. This pulse sequence is then repeated, with a 20μs delay between sequences to allow the emitter to reach thermal equilibrium before the next sequence commences. The procedure is repeated for increasing values of $\tau$ until the peak value of the first peak P1 and the second peak P2 are the same, suggesting that the emitter has reached thermal equilibrium at this time delay. From this data, we extracted the ratio between the intensity of the P1 and P2 as a function of $\tau$ (Fig. 5c) and fitted it with the expression:

$$1 - A * \exp\left(-\frac{\tau}{t_A}\right) - B * \exp\left(-\frac{\tau}{t_B}\right), \quad (3)$$

where time constants $t_A = 0.47 \pm 0.06$ μs and $t_B = 4.0 \pm 0.6$μs.

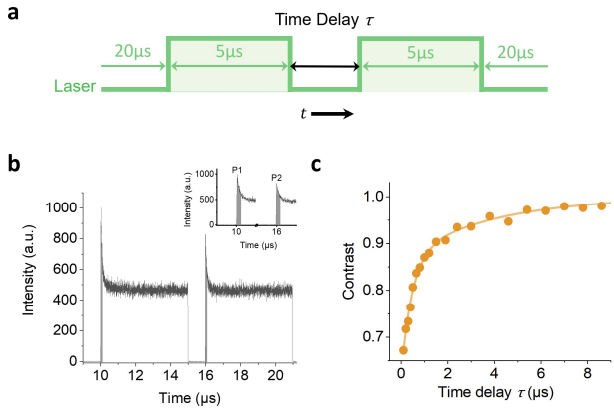

**Figure 5 | Laser pulse measurement. a**, Laser pulse sequence with a time delay $\tau$ in between pulses. Each laser pulse is switched on for 5μs, and a 20μs recovery time between each set of two laser pulses is implemented. **b**, PL time trace of the laser sequence in (a) at a time delay of $\tau = 1$ μs. The areas under the curve shaded in grey were integrated. The inset shows a close-up image of the two PL peaks. **c**, The ratio of the integrated areas (P2/P1) as a function of time delay. The orange scatter plot is the experimental points, while the solid line is the fitted curve.

These time constants should reflect the decay process to the ground state $|1\rangle$. We exclude the decay process from the emitting state $|2\rangle$ since it is significantly faster than these time constants (lifetime ~0.1ns based on $g^{(2)}$ measurement in Fig. 2b) and consistent with previous time-resolved PL measurements on telecommunication SPE [28,29]. Hence, these time constants are likely related to the lifetime of the metastable states $|3\rangle$ and $|4\rangle$ (i.e., the inverse of the decay rates $k_{41}$ and $k_{31}$ in Fig. 2a). This result supports the existence of at least two metastable states, consistent with the ODMR results and supporting the fact that the spin-dependent ODMR mechanism originates at the metastable state, similar to the ST1 defect in diamond [42]. The low CW-ODMR contrast and the lack of pulsed ODMR could also be attributed to the short lifetime of these states. Interestingly, the dynamics with time scales $t_A$ and $t_B$ are absent in the fitting of $g^{(2)}_{raw}(\tau)$ measured at the same laser power under CW mode, which could indicate optically driven de-trapping of the metastable states $|3\rangle$ and $|4\rangle$ under the same excitation laser. Further analysis and modelling of the SPE supporting this hypothesis can be found in Note 1 and Fig. S1 in [37].

In conclusion, we have successfully demonstrated the existence of telecommunication SPEs in GaN that exhibit ODMR under ambient conditions. The detailed characterization of the magneto-optical properties and optical dynamics of these SPEs, including zero-field splitting, the gyromagnetic ratio, and the metastable states' lifetime, gives us a better understanding of the origin of the telecom SPEs and sets the foundation to improve the performance of the spin defect. Additionally, the determination of the spin quantization axis orientation with respect to the GaN lattice provides necessary quantitative information to benchmark potential defect candidates accounting for the single-photon emission at telecommunication wavelengths in GaN. Our study not only expands the available and scarce spin-defects operating in the telecommunication range, but also opens new avenues for research into solid-state defects and their applications in quantum sciences for real-field applications at room-temperature.

### Acknowledgements

This work is supported by Singapore Quantum Engineering Program (NRF2021-QEP2-01-P02, NRF2022 QEP2-02-P13), Singapore National Research foundation (NRF-CRP22 2019-0004, NRF2023-ITC004-001), Singapore Ministry of Education Tier 2 Grant (MOE-T2EP50221-0009, MOE-T2EP50222-0018), Agency of Science, Technology, and Research (A*STAR), Singapore, MTC IRG (Project No. M21K2c0116), and by the French government through the National Research Agency (ANR) in the context of the Plan France 2030 through project reference ANR-22-PETQ-0011, and by the ANR through project ANR-22-CE47-0006-01.

### Methods

*Optical Characterisation Setup*

Both the excitation and collection were performed through the same microscope objective (OLYMPUS, LCPLN 100× IR) with an NA of 0.85. The excitation laser is a 980nm CW laser diode filtered by a 1000nm shortpass filter (Thorlabs, FESH1000), before entering the 1000nm longpass dichroic mirror beam splitter (Thorlabs, DMLP1000R). The collection signal is filtered with various longpass, shortpass, and bandpass filters from Thorlabs and Edmund Optics to isolate the spectral signal of the SPE and remove as much background signal as possible. The stage to mount the GaN sample is a 3D nanopositioner (Physik Instrumente, P-611.3S).

*Sample Fabrication*

We used a 2μm thick GaN thin films grown by MOCVD using a close-coupled showerhead Thomas Swann reactor on a *c*-plane sapphire substrate. The growth employed a SiN surface treatment to induce a 3D GaN nucleation in order to reduce the density of dislocations and a low temperature GaN buffer layer. Precursors used were trimethylgallium (TMGa) and ammonia (NH3). The microwave strip lines were fabricated by electron-beam lithography and consisted of 8 and 200 nm of Cr and Au films, respectively.

*ODMR Setup*

A DAQ (National Instruments USB-6343-BNC) reads the photon counts during the read-out window and norm window, which are gated by the pulse generator (Swabian Instruments Pulse Streamer 8/2). The microwave signal was generated by a signal generator (Stanford Research Systems SG386) and then amplified by a microwave amplifier (Mini circuits ZHL-30w-252-s+, ZHL-16w-43-s+, TVA-4w-422a+) before being delivered to the sample via a lithographically deposited strip line on the top surface of the GaN sample. The magnetic field was applied by a N52 permanent magnet and mounted on a motorized linear stage (Thorlabs, MT3-Z9).

# Supplementary Material for

# Room-temperature Optically Detected Magnetic Resonance of Telecom Single Photon Emitters in GaN

John J.H. Eng [1,2†], Zhengzhi Jiang [3, 4†], Max Meunier [1], Abdullah Rasmita [1], Haoran Zhang [1], Yuzhe Yang [1], Feifei Zhou [1], Hongbing Cai [1,5], Zhaogang Dong [2], Jesús Zúñiga Pérez [1, 6, 7*], Weibo Gao [1, 5*]

*Corresponding Author. Email: jesus.zuniga@ntu.edu.sg, wbgao@ntu.edu.sg

**Supplementary Note 1: Theoretical Model for the SPE**

The $g^{(2)}_{raw}(\tau)$ experimental data is first treated with a background correction factor to ensure that the theoretical model could represent the SPE model more accurately without the influence of background emission. The background corrected $g^{(2)}(\tau)$ is defined as: $g^{(2)}(\tau) = [g^{(2)}_{raw}(\tau) - (1 - \rho^2)]/\rho^2$ where $\rho = s/(s + b)$ and $s$ is the signal's intensity while $b$ is the background emission intensity [35,43-45]. We then extract the fitting parameters ($t_1$ and $t_2$) using the general empirical model (main text Eq. 1 and Table S1 for the fitting parameters) and taking into account the $t_A$ and $t_B$ characteristic times measured owing to the pump-probe (or time-gated) experiment (main text Fig. 5). At 1mW, the fitting parameters $t_1$ and $t_2$ do not reproduce the $t_A$ and $t_B$ times despite being measured at the same optical excitation power. This suggest that there are (1) two decay channels via the metastable states with decay times $t_A$ and $t_B$, and (2) optically assisted de-trapping mechanisms that bypass the decay channels via these metastable states under constant illumination, since $t_A$ and $t_B$ are longer than $t_1$ and $t_2$. Hence, we propose a 4-level energy system model (main text Fig. 2a) with two metastable triplet states $|3\rangle$ and $|4\rangle$ characterized by de-trapping rates $k_{32}$ and $k_{42}$, respectively. It is worth mentioning that a more accurate model would describe the de-trapping of triplet states $|3\rangle$ and $|4\rangle$ via optical pumping into a higher intermediate triplet state, before decaying back to $|2\rangle$ via intersystem crossing. However, for simplicity, we treat this process as one effective transition to minimize the number of fitting parameters.

To verify the viability of the model, we fit the background corrected $g^{(2)}(\tau)$ by solving the coupled rate equations as described pictorially in Fig. 2a of the main text. The relationship between $g^{(2)}(\tau)$ of the SPE and the state population is defined as: $g^{(2)}(t) = n_2(t)/n_2(t \to \infty)$ where $n_x(t)$ is the evolution of the population of state $|x\rangle$ in time [35]. We limit the decay rate of $k_{31}$ and $k_{41}$ to $1/t_A$ and $1/t_B$ respectively, constrained by the fitted uncertainty shown in Fig. 5c. The remaining transition rates (except $k_{12}$, $k_{32}$, and $k_{42}$) are assumed to be unaffected by optical excitation power and used as shared fitting parameters that are unbounded positive values (i.e. $[0, inf]$). The rates $k_{12}$, $k_{32}$, and $k_{42}$ are also unbounded positive values but are not shared, as they should be power dependent. The resulting fit of the $g^{(2)}(t)$ at several optical excitation powers is shown in Figure S1a. During the $g^{(2)}(\tau)$ measurement, since the emitter is under CW excitation, the de-trapping of the metastable states would be active and should be faster than the decay to the ground state, which could explain the absence of $t_A$ and $t_B$ in the fitting with the general empirical model. Furthermore, the resulting fitted rates $k_{12}$, $k_{42}$, and $k_{32}$ show a linear dependence with optical excitation power (Fig. S1b), further supporting the model's viability.

Interestingly, the rates $k_{23} = 45 \pm 31$MHz and $k_{24} = 35 \pm 25$MHz (decay to the metastable states) are faster than $k_{31} = 2.1 \pm 0.2$MHz and $k_{41} = 0.25 \pm 0.03$MHz (decay from metastable states), similar to the case of NV⁻ in diamond [46]. One possible explanation for this could be that the excited state is higher in energy than the metastable states, which typically results in faster non-radiative transition rate facilitated by spin-orbit coupling and phonon interactions.

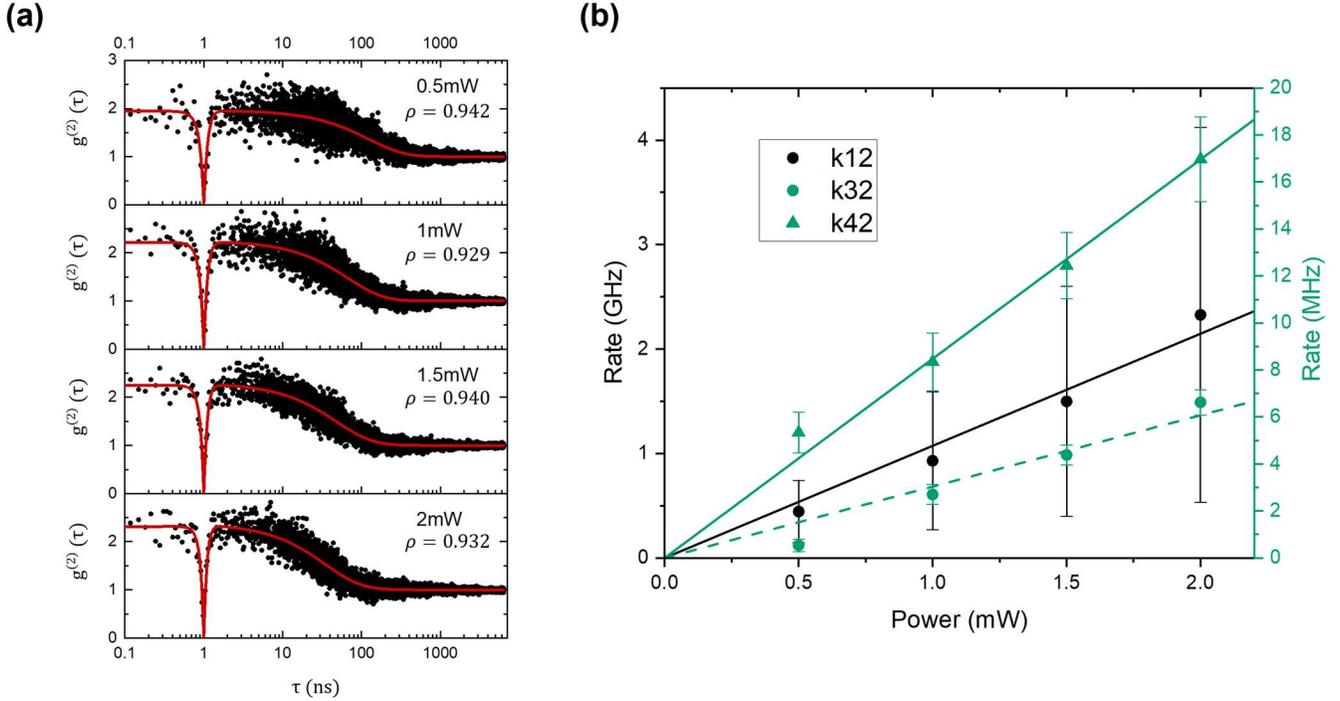

**Figure S1 | Fitting of background correct $g^{(2)}(\tau)$ using the proposed model. a,** Background corrected $g^{(2)}(\tau)$ (black scatter plot) offset by $\tau = 1$ measured at various optical excitation powers fitted by solving the coupled rate equations (red solid line). **b,** Optical excitation power dependent rates. The plots in black correspond to the left black y-axis in GHz, while the plots in green correspond to the right green y-axis in MHz. The scatter plots are the fitting parameters $k_{12}$ (black circles), $k_{32}$ (green circles), and $k_{42}$ (green triangles). The error bars for each point are the 95% confidence interval resulting from the fitting. The solid black and green line as well as the dashed green line are the fitting lines for the scatter plots of $k_{12}$, $k_{42}$, and $k_{32}$ respectively.

| Laser power | 0.5mW | 1mW | 1.5mW | 2mW |
|---|---|---|---|---|
| $C_1$ | 1.53 ± 0.13 | 1.78 ±0.12 | 1.68 ± 0.10 | 1.74 ± 0.10 |
| $t_1$ (ns) | 0.13 ± 0.014 | 0.11 ± 0.01 | 0.12 ± 0.01 | 0.11 ± 0.01 |
| $C_2$ | 0.846 ±0.004 | 1.058 ± 0.004 | 1.110 ± 0.004 | 1.151 ± 0.004 |
| $t_2$ (ns) | 128.0 ± 1.4 | 70.0 ± 0.6 | 50.7 ± 0.4 | 37.1 ± 0.3 |
| $g^{(2)}_{raw}(0)$ | 0.32 ± 0.14 | 0.27 ± 0.12 | 0.43 ± 0.10 | 0.42 ± 0.10 |

**Table S1 | Fitting parameters for $g^{(2)}_{raw}(\tau)$ measured at optical excitation power of 0.5, 1, 1.5, and 2mW** (The fitting function for $g^{(2)}(\tau)$ is given in Eq. (1) in the main text.

**Supplementary Note 2: Alignment and Calibration of Magnetic Field.**

In previous works, we demonstrated that the photoluminescence (PL) signal from boron vacancy defects in hBN changes in response to variations in the magnetic field [47]. This property is utilized to precisely align the z-axis magnetic field. The alignment method is outlined below and employs a local optimization strategy described in the following steps:

1. The initial step involves scanning the z position of the magnet to identify the magnetic field strength at which the excited-state level anti-crossing (ESLAC) occurs for the boron vacancy in hBN, as depicted in the shaded area in Fig. S2(a).
2. After determining the z position that results in ESLAC, the x and y positions of the magnet are scanned to ascertain the coordinates that yield the highest fluorescence intensity, as shown in Fig. S2(b). Positioning the magnet at these (x, y) coordinates ensures that the magnetic field it generates is aligned along the z-axis of the sample, with the alignment error being less than 0.5 degrees.

After aligning the magnetic field, we proceeded to calibrate its strength using continuous-wave optically detected magnetic resonance (CW-ODMR) of boron vacancies. For each z position, we conducted CW-ODMR measurements on boron vacancies and applied a Gaussian peak function to fit the data, which indicates the magnetic field values. Subsequently, we mapped the position-to-field data using an exponential function for a more accurate representation [48].

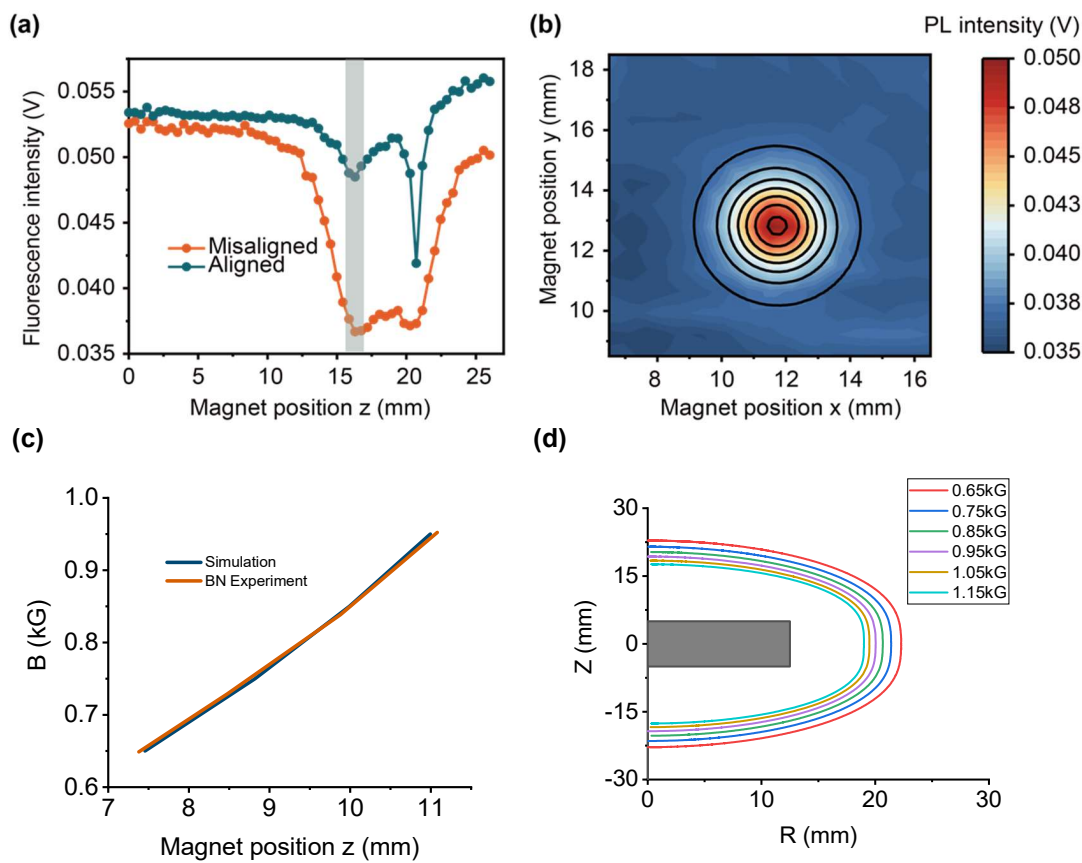

**Figure S2 | Magnetic field alignment process. a,** Variation in fluorescence intensity corresponding to changes in the magnet's z position. The shaded region highlights the range indicating where ESLAC is achieved. **b,** Results from scanning across the XY plane while maintaining the z position at the point of ESLAC occurrence. **c,** Comparison of magnetic flux density as a function of magnet's stage position calibrated by the hBN sample with the simulation results (COMSOL). **d,** Simulation results of the spatial distribution of various magnetic flux density B values. The gray rectangle is the magnet. Any point along the same-colored solid line has the same magnetic flux density magnitude but differing vector direction.

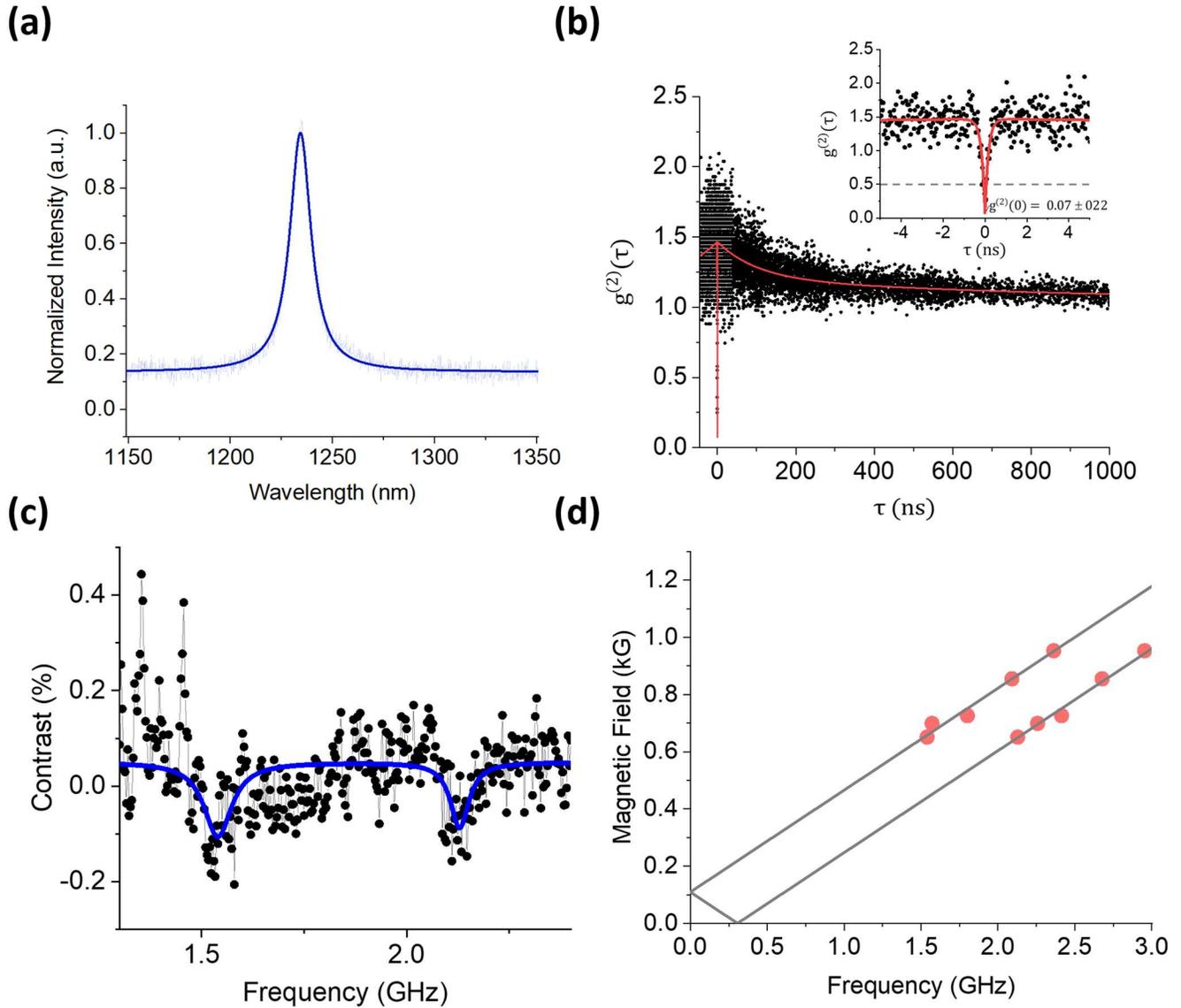

**Figure S3 | Magnetic and optical properties of another SPE (SPE #2). a**, PL spectrum measured at 1mW. **b**, Second-order autocorrelation $g_{raw}^{(2)}(\tau)$ measured at 0.5mW with a $g_{raw}^{(2)}(\tau) = 0.07 \pm 0.22$ after fitting. The black scatter plots and red solid lines are the experimental points and fitting lines, respectively. The experimental data is fitted with Eq. (1) in the main text. **c**, CW-ODMR spectrum measured at a microwave power of 40 dBm. The black scatter plot and blue solid line are the experimental and fitting line, respectively. The scatter plot is fitted with two Lorentzian peaks. **d**, Magnetic field dependence as a function of CW-ODMR resonant frequency. The red scatter plot and gray solid line are the experimental points and fitted line, respectively. The experimental points are fitted with the same expression $\left| D \pm \sqrt{E^2 + (\gamma B_z)^2} \right|$ as in the main text. The axial zero field splitting $D = 0.307$ GHz while the transverse zero field splitting $E$ is zero. The gyromagnetic ratio $\gamma$ is 2.81GHz while the field is aligned with the $c$-axis of the GaN.